\documentclass[twocolumn,showpacs,preprintnumbers,amsmath,amssymb,superscriptaddress]{revtex4-1}

\usepackage{graphicx}
\usepackage{dcolumn}
\usepackage{bm}
\usepackage{tabularx}
\usepackage{ulem}
\newcolumntype{M}{>{\centering\arraybackslash}m{1.85cm}}
\usepackage[export]{adjustbox}
\usepackage{float}
\usepackage{color}   
\usepackage{hyperref}
\hypersetup{
    colorlinks=true, 
    linktoc=all,     
    linkcolor=blue,  
}

\makeatletter
\newcommand{\colorcaption}[2][]{%
  \begingroup%
  \renewcommand{\@caption@fignum@sep}{ (Color online). }%
  \caption[#1]{#2}%
  \endgroup%
}
\makeatother
\newcommand\T{\rule{0pt}{3ex}}       
\newcommand\B{\rule[-1.5ex]{0pt}{0pt}} 

\begin{document}

\title{Shell-model description for the properties of the forbidden $\beta^-$ decay in the region ``north-east" of $^{208}$Pb }

\author{ Shweta Sharma}
\address{Department of Physics, Indian Institute of Technology Roorkee, Roorkee 247667, India}
\author{Praveen C. Srivastava}
\email{Corresponding author: praveen.srivastava@ph.iitr.ac.in}
\address{Department of Physics, Indian Institute of Technology Roorkee, Roorkee 247667, India}
\author{ Anil Kumar}
\address{Department of Physics, Indian Institute of Technology Roorkee, Roorkee 247667, India}
\author{Toshio Suzuki}
\address{Department of Physics, College of Humanities and Sciences, Nihon University, Sakurajosui 3, Setagaya-ku, Tokyo 156-8550, Japan}

\date{\hfill \today}
\begin{abstract}

In the present work, we report a comprehensive shell-model study of the $\log ft$ values for the forbidden $\beta^-$ decay transitions in the north-east region of $^{208}$Pb. For this we have considered  $^{210-215}$Pb $\rightarrow$ $^{210-215}$Bi and $^{210-215}$Bi $\rightarrow$ $^{210-215}$Po transitions. We have performed shell-model calculation using KHPE interaction 
in valence shell 82-126 for protons and 126-184 for neutrons without any truncation. We have also calculated half-lives and  Q-values for the concerned nuclei.  Recently several $\log ft$ values are observed corresponding to $\beta^-$ decay from (8$^-$) isomeric state  of $^{214}$Bi$^m$ at CERN-ISOLDE facility [Phys. Rev. C {\bf 104}, 054301 (2021)], and for the first time we have reported shell-model results for these transitions.
 
\end{abstract}

\maketitle

\section{Introduction}
 The $r$-process plays a significant role in the nucleosynthesis of heavier nuclei in astrophysics \cite{cowan91}. It is believed to occur in core-collapse supernovae \cite{core} or neutron star mergers\cite{merger}, but the actual site of $r$-process nucleosynthesis is still an open area for investigation \cite{rprocess}. The main conditions required for the $r$-process nucleosynthesis \cite{Burbidge57} are high temperature and large neutron density \cite{cameron57,woos94}. Further, the abundance pattern of $r$-process nuclei shows enhanced peaks near neutron shell closures. $^{208}$Pb is the heaviest doubly magic stable nucleus with 82 protons and 126 neutrons. Due to the increased stability of doubly magic nuclei compared to the single magic nuclei and others, the region around these nuclei has always been an area of great interest for investigation. The measurement of half-lives of nuclei near N = 126 is difficult. Thus, it is highly desirable to give theoretical predictions for half-lives around this region. Suzuki \textit{et al.} \cite{suz12} have evaluated beta-decay properties of $r$-process nuclei near N=126 isotones using shell-model calculations.  First forbidden beta-decay competes with allowed Gamow-Teller and Fermi beta-decay in the region of N=126 $r$-process nuclei \cite{car,bru21}.  It is thus crucial to study these nuclei in this region for a better understanding of the abundance pattern of these $r$-process nuclei.

 
 Beta-decay can be divided into two categories: allowed and forbidden, based on the value of the angular momentum of the emitted leptons. Allowed transitions correspond to the no change in parity and $l = 0$ state of the emitted leptons relative to the nucleus, while $l>0$ corresponds to the forbidden transitions. Further, based on the value of spin angular momentum, these transitions can be characterized in two categories, Fermi and GT transitions. Fermi transitions correspond to spin non-flip with $S=0$, whereas GT transitions correspond to  spin-flip with $S = 1$. Further, forbidden transitions can be divided into the unique forbidden beta-decay and the non-unique forbidden beta-decay. In unique forbidden beta-decay, $\Delta J = l+1$, while in non-unique forbidden beta-decay, $\Delta J = l-1, l $, where $l$ is the degree of forbiddenness. Change in parity for forbidden beta-decay is positive (even) for an even degree of forbiddenness, while negative (odd) for an odd degree of forbiddenness. Beta- decay properties have been evaluated using effective values of the weak coupling constants in Refs. \cite{joel12017,joel22017}. Recently, using shell-model, our group has calculated first-forbidden beta-decay properties of $^{207}$Hg $\rightarrow $ $^{207}$Tl in Ref. \cite{anil21npa}.
 
 There are several approaches that can be used to calculate beta-decay properties. First is the macroscopic approach, i.e., Gross theory of beta-decay \cite{gross} and second is global semi-microscopic approaches such as quasi-particle random phase approximation (QRPA) \cite{borzov2000}, density functional theory (DFT) \cite{nom20}, Hartree-Fock-Bogoliubov method (HFB) \cite{engel99}, etc. These models underestimate the residual interaction between nucleons, which reduces the Gamow-Teller (GT) strengths towards lower excitation energies \cite{Cuenca,martinez03,martinez99}. The third approach is the microscopic approach, i.e., shell-model.  We are using large-scale shell-model in the study of beta-decay properties. Further, in shell-model, quenching factors for the weak axial and vector coupling constants are needed to reproduce reliable data \cite{jouni18,anil20}. Several efforts have been done to calculate  quenching factors in Pb region: Warburton \cite{warburton90} found that the quenching factors for the axial and vector coupling constants are different and it comes out to be ($g_{\rm A}/g_{\rm A}^{\rm free}$, $g_{\rm V}/g_{\rm V}^{\rm free}$)=(0.47, 0.64) in the Pb region. Further, the mesonic enhancement factor has been calculated in Refs. \cite{enhancement,warburton91} for the first forbidden beta-decay with $\Delta J ^{\pi} = 0^{-}$ for A = 205-212, where the enhancement factor is defined as the proportion of the axial-charge matrix element $\gamma_5$ in first-forbidden beta-decay to its impulse-approximation value. This value comes out to be $\epsilon_{\text{MEC}}$ = 1+$\delta_{MEC}$ = 2.01$\pm$0.05. Later on, Rydstrom \textit{et al.} in Ref. \cite{ryd90} found out two sets of quenching factors in the Pb region: first one being ($g_{\rm A}/g_{\rm A}^{\rm free}$, $g_{\rm V}/g_{\rm V}^{\rm free}$)=(0.34, 0.67) and the second one was ($g_{\rm A}/g_{\rm A}^{\rm free}$, $g_{\rm V}/g_{\rm V}^{\rm free}$)=(0.51, 0.30). Furthermore, Zhi \textit{et al.} \cite{zhi} performed large-scale shell-model calculations for $r$-process waiting point nuclei with N = 50, 82, 126, including both GT and first forbidden transitions. They found that the shell-model  overestimate the transition strengths in the GT and forbidden beta-decays. Thus,  they found the quenching factors i.e. ($g_{\rm A}/g_{\rm A}^{\rm free}$, $g_{\rm V}/g_{\rm V}^{\rm free}$)=(0.38, 0.51). 

 In the present work, the $\log ft$ values, average shape factor, and half-lives have been calculated for $^{210-215}$Pb $\rightarrow$  $^{210-215}$Bi and $^{210-215}$Bi $\rightarrow$ $^{210-215}$Po transitions and compared with the available experimental data. These beta-decay properties have been computed using two sets of quenching factors: the first one being calculated from our work using the chi-squared fitting method, whereas the second one is taken from Ref. \cite{zhi}. To the best of our knowledge,
  theoretical estimates for these nuclei mentioned are carried out for the first time except for $\log ft$ values for $^{211}$Pb to $^{211}$Bi transitions \cite{enhancement}, which were calculated by using old experimental data with truncated model
 space using KHPE interaction \cite{war91}. In our work, we have performed large-scale shell-model calculations without using any truncation. Since $\beta$-decay is very sensitive to the Q-value, so, it is crucial to use a precise Q-value. Thus, we have also calculated Q-value using shell-model calculations and used  them in the calculation of beta-decay properties. The shell-model results have also been calculated where experimental data is  not available. Based on our calculated $\log ft$ values, we have confirmed spin and parity of several states where experiments were unable to make unique assignments.
 
 Recently, a new experiment has been performed at the CERN-ISOLDE facility to study $^{214}$Bi isotope using $\gamma$ ray spectroscopy, and a new isomeric state $(8^-)$ has been identified. Further, beta-decay of this $^{214}$Bi$^m$ isomer to various spin parity states of $^{214}$Po was studied, and its $\log ft$ values and beta-decay feeding  fractions have been reported. Moreover, the experimental energy spectra of $^{214}$Po and yrast and yrare states of $^{214}$Bi were compared with the KHPE and H208 interactions, but $\log ft$ values using these shell-model interactions were not calculated. Thus, for the first time, these $\log ft$ and average shape factor values  have been calculated within the framework of shell-model and compared with the experimental values according to the decay mode as presented in Ref. \cite{iso}.
 
 The content of this paper is organized as follows: Section II depicts theoretical formalism in which shell-model Hamiltonian, beta-decay theory, and quenching factor are briefly discussed. Further, the calculation of the quenching factor with the help of the chi-squared fitting method is given in this section.  In Section III,   $\log ft$ values, average shape factors, and half-lives for the concerned nuclei are evaluated, and computations of $Q$-values are also carried out.  Finally, conclusion is given in Section IV.    

\section{Formalism}
\subsection{Shell-model Hamiltonian}
The nuclear shell-model Hamiltonian can be expressed as combination of a single-particle energy term and a two-nucleon interaction term \cite{jouni}. The shell-model Hamiltonian has the form
\begin{equation}
H = T + V = \sum_{\alpha}{\epsilon}_{\alpha} c^{\dagger}_{\alpha} c_{\alpha} + \frac{1}{4} \sum_{\alpha\beta \gamma \delta}v_{\alpha \beta \gamma \delta} c^{\dagger}_{\alpha} c^{\dagger}_{\beta} c_{\delta} c_{\gamma},
\end{equation}
where $\alpha = \{n,l,j,t\}$ stands for single-particle state and the corresponding single particle energy is denoted by $\epsilon_{\alpha}$. $c^{\dagger}_{\alpha}$ and $c_{\alpha}$ stands for creation and annihilation operators. 
$v_{\alpha \beta \gamma \delta} = \langle\alpha \beta | V | \gamma \delta\rangle $ are the antisymmetrized two-body matrix elements.

\subsection{$\beta$ decay theory for allowed and forbidden transitions} \label{beta}
The theoretical formalism of beta-decay theory is briefly explained here.  Here, we will give brief details of formalism about allowed and forbidden beta-decay. 
One can find more detailed formalism in Refs. \cite{hfs,beh82}.
This formalism is based on impulse approximation \cite{jouni} i.e., the decaying nucleon does not feel strong interaction with the remaining nucleons and only feels weak interaction at the instance of decay. Here, the remaining nucleons will act as spectator. The total half-life is the inverse of the decay rate, and it can be defined as
\begin{equation}
\frac{1}{T_{1/2}}=\sum_k \frac{1}{t_{1/2}^{(k)}},
\end{equation}
where $t_{1/2}^{(k)}$ is the partial half-life to the final state $k$. 
The partial half-life is related to transition probability as
\begin{eqnarray}\label{hf1}
t_{1/2}=\frac{\text{ln}(2)}{\int_{m_ec^2}^{W_0}{P(W_e)dW_e}},
\end{eqnarray}
 where the integrand in the denominator is the transition probability, and $m_e$ is the electron mass.
The probability of the emitted beta particle to have energy between $W_e$ and $W_e+dW_e$ has the form
\begin{equation} \label{eq1}
\begin{split}
P(W_e)dW_e & = \frac{G_\text{F}^2}{(\hbar{c})^6}\frac{1}{2\pi^3\hbar}C(W_e)p_ecW_e(W_0-W_e)^2 \\ & \times{F_0(Z,W_e)dW_e},
\end{split}
\end{equation}
where $G_\text{F}$ is the effective coupling constant, i.e., the Fermi coupling constant determines the strength of beta interaction, $p_e$ and $W_e$ are the momentum and energy of the emitted beta particle, respectively. $W_0$ is the endpoint energy, i.e., the maximum energy attained by the emitted beta particle in the beta-decay process. The $C(W_e)$ is the shape factor that depends on electron energy, and $F_0(Z,W_e)$ is the Fermi function included in the expression for Coulomb interaction between the beta particle and remaining nucleus. For the simplification of the integration, dimensionless quantities are introduced such as $w_0=W_0/m_ec^2$, $w_e=W_e/m_ec^2$, and $p=p_ec/m_ec^2=\sqrt{(w_e^2-1)}$. Thus, the dimensionless integrated shape function has the form

\begin{eqnarray} \label{tc}
f=\int_1^{w_0}C(w_e)pw_e(w_0-w_e)^2F_0(Z,w_e)dw_e.
\end{eqnarray}
 
 The shape factor does not depends on electron energy in case of allowed transition i.e. $C(w_e)=B(GT)$, where $B(GT)$ is the Gamow-Teller transition probability. Thus,
 
 \begin{eqnarray}\label{mgt}
C(w_e)=\frac{g_A^2}{2J_i+1}|\mathcal{M}_{\text{GT}}|^2, 
\end{eqnarray}
where the $J_i$ is the initial angular momentum and $g_A$ is the axial-vector coupling constant, and the $\mathcal{M}_\text{GT}$ stands for the Gamow-Teller nuclear matrix element \cite{Brown_Wildenthal}.

 Thus, the phase-space factor becomes
 
 \begin{equation}
	f_{0}= \int_1^{w_0}pw_e(w_0-w_e)^2F_0(Z,w_e)dw_e.
\end{equation}

For forbidden beta-decay, the shape factor is given by

\begin{eqnarray} \label{eq2}
\begin{split}
C(w_e)  = \sum_{k_e,k_\nu,K}\lambda_{k_e} \Big[M_K(k_e,k_\nu)^2+m_K(k_e,k_\nu)^2 \\
-\frac{2\gamma_{k_e}}{k_ew_e}M_K(k_e,k_\nu)m_K(k_e,k_\nu)\Big],
\end{split}
\end{eqnarray}
where $K$ is the forbiddenness order and $k_e$ and $k_\nu$ are the positive integers emerging from partial wave expansion of the leptonic wave function. The quantities $M_K(k_e,k_\nu)$ and $m_K(k_e,k_\nu)$ are expressed in terms of the nuclear matrix elements (NMEs) containing nuclear structure information and the leptonic phase space factors. The auxiliary quantities $\gamma_{k_e}$ and $\lambda_{k_e}$ can be written as 
\begin{align*}
	\gamma_{k_e} & =\sqrt{k_e^2-(\alpha{Z})^2}&\mbox{and} \,\,\,\, \lambda_{k_e} & ={F_{k_e-1}(Z,w_e)}/{F_0(Z,w_e)},
\end{align*}
where $\lambda_{k_e}$ stands for Coulomb function and $F_{k_e-1}(Z,w_e)$ is the generalized Fermi function \cite{mika2017,must2006} which has the form
\begin{eqnarray} \label{eq3}
F_{k_e-1}(Z,w_e) &=4^{k_e-1}(2k_e)(k_e+\gamma_{k_e})[(2k_e-1)!!]^2e^{\pi{y}} \nonumber \\
 & \times\left(\frac{2p_eR}{\hbar}\right)^{2(\gamma_{k_e}-k_e)}\left(\frac{|\Gamma(\gamma_{k_e}+iy)|}{\Gamma(1+2\gamma_{k_e})}\right)^2.
\end{eqnarray}

The auxilliary quantity $y=(\alpha{Zw_e}/p_ec)$, where $\alpha=1/137$ is the fine structure constant.

The NMEs \cite{Anil,anil21} can be described as

\begin{align}
\begin{split}
^{V/A}\mathcal{M}_{KLS}^{(N)}(pn)(k_e,m,n,\rho)& \\ =\frac{\sqrt{4\pi}}{\widehat{J}_i}
\sum_{pn} \, ^{V/A}m_{KLS}^{(N)}(pn)(&k_e,m,n,\rho)(\Psi_f\parallel [c_p^{\dagger}\tilde{c}_n]_K\parallel \Psi_i),
\label{eq:ME}
\end{split}
\end{align}
where ${\widehat{J}_i}=\sqrt{2J_i+1}$ with $J_i$ being the initial angular momentum and the summation runs over protons and neutrons single particle states. The quantity ${^{V/A}m_{KLS}^{(N)}}(pn)(k_e,m,n,\rho)$ stands for the single particle matrix elements (SPMEs) which is independent of he choice of nuclear models. $(\Psi_f|| [c_p^{\dagger}\tilde{c}_n]_K || \Psi_i)$ is the one body transition density (OBTDs) which vary for different nuclear models. Here, $(\Psi_i)$ and $(\Psi_f)$ are the initial and final nuclear states. The SPMEs are calculated with the help of formalism given in Ref. \cite{beh71} whereas the OBTDs are calculated from the shell-model using NuShellX \cite{nushellx} and KSHELL \cite{shim19}.
The partial half-life is usually expressed as comparative half-life or the reduced half-life, which is given as

\begin{equation}
	ft_{1/2}=\kappa,
\end{equation}
 where $\kappa$ is constant value which is expressed as \cite{patri}
 
\begin{eqnarray}
\kappa=\frac{2\pi^3\hbar^7\text{ln(2)}}{m_e^5c^4(G_\text{F}\text{Cos}\theta_\text{C})^2}=6289~\mathrm{s},
\end{eqnarray}
where $\theta_\text{C}$ is the Cabibbo angle which is the mixing angle between two generation of quarks.

 Usually, $ft$ are expressed in terms of 'log $ft$ values' because $ft$ values are large. Thus,

\begin{equation*}
	\mbox{log} ft \equiv \mbox{log}(f_{0}t_{1/2}).
\end{equation*}

The phase space factor is sensitive to the Q-value. Therefore, it is essential to evaluate it precisely. Thus, the Q-value \cite{VikasK} can be expressed as

\begin{equation}
	Q(\beta^-)=E_{g.s.}^{par}-E_{g.s.}^{dau}+\delta m,
\end{equation}
where $\delta m=(m_n-m_p-m_e)c^2=0.78$ MeV. $E_{g.s.}^{par}$ and $E_{g.s.}^{dau}$ stands for ground state binding energy for the parent and daughter nuclei, respectively.
The binding energy of the ground state is given by

\begin{equation}
	E=E_{SM}+E_{core}+E_C(Z,N).
\end{equation}
Here $E_{SM}$ is the shell-model calculated binding energy and $E_{core}$ is the binding energy of the core considered, and $E_{C}(Z, N)$ is the Coulomb energy which can be calculated from the formalism given in Refs. \cite{duf95,caurier99}.




\begin{figure}
  \includegraphics[width=85mm]{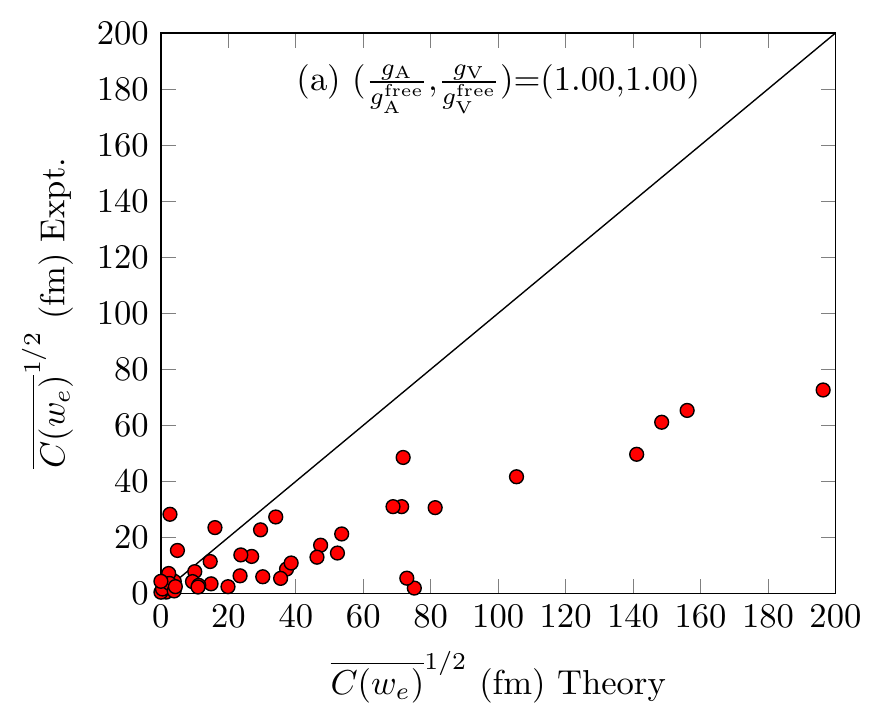}
   \includegraphics[width=85mm]{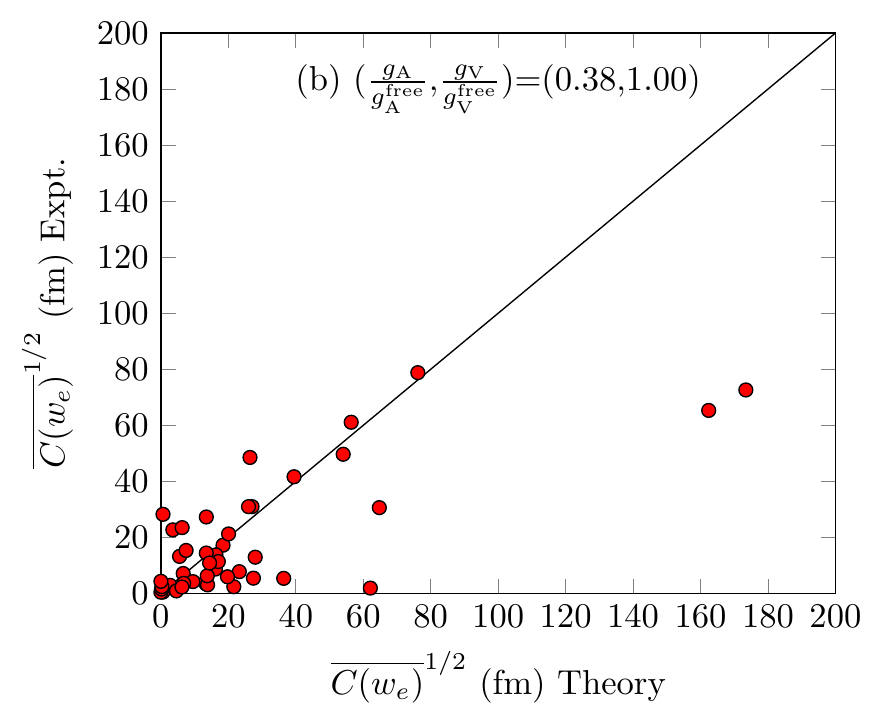}
   \includegraphics[width=85mm]{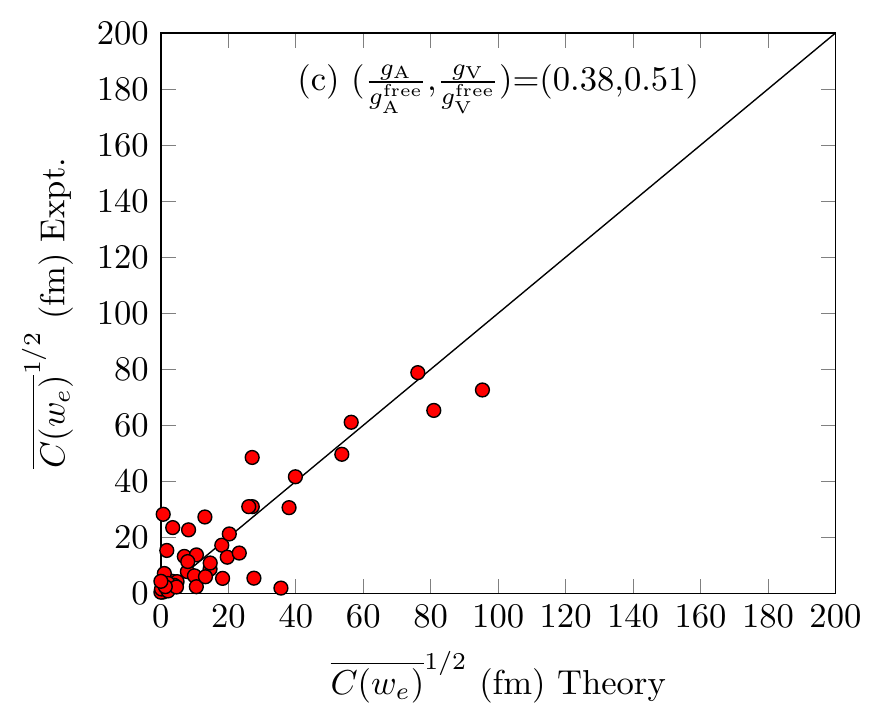}
  \label{fig:occupancy}
\caption{\label{occupancy}Comparison of calculated and experimental average
 shape factors for allowed and first-forbidden transitions by using different values of quenching factors for axial and vector coupling constants. }
\end{figure}

\subsection{Quenching factor}

In beta-decay, the Gamow-Teller and forbidden strengths get overestimated in the shell-model calculations.  The weak coupling constants $g_{\rm V}$ and $g_{\rm A}$ are included in these Gamow-Teller and forbidden strengths, where $g_{\rm V}$ is the vector coupling constant decided by CVC (Conserved Vector Current) theory and $g_{\rm A}$ is the axial-vector coupling constant determined by PCAC (Partial Axial Vector Current) theory. The free nucleon values of these weak coupling constants are $g_{\rm V}$=1.0 and $g_{\rm A}$=1.27. These values get affected by many nucleon correlations like model space truncation in shell-model calculations and other nuclear medium effects. Therefore, the values of weak coupling constants get heavily quenched in the heavier mass region. Thus, we use effective values of these weak coupling constants.
According to Behrens and B$\ddot{\text{u}}$hring \cite{beh82}, the average shape factor is given by

\begin{equation}
	\overline{C(w_{e})} = f/f_{0}\,.
\end{equation}

 Here $f$ is the phase space factor and $f_0$ is the phase space factor for allowed transitions. For allowed GT transitions, this shape factor is independent of electron energy. 
 However, for n$^{\text{th}}$ forbidden transition, the average shape factor \cite{enhancement} comes out to be 

\begin{equation}
	\overline{C(w_{e})} (fm^{2n}) =\frac{6289 \lambdabar_{\text{Ce}}^{2n}}{ft}.
\end{equation} 

Thus, for the first forbidden transition, the average shape factor \cite{pri21} has the form

\begin{equation}
	\overline{C(w_{e})} (fm^2) =\frac{6289 \lambdabar_{\text{Ce}}^2}{ft} = \frac{9378 \times 10^5}{ft},
\end{equation} 
where $ \lambdabar_{\text{Ce}} $ is the reduced Compton wavelength of electron. Now, for allowed GT transitions, the operator is just $\bm{\sigma} \bm{\tau}$, whereas there are six Nuclear Matrix Elements (NMEs) for the first forbidden transition \cite{jouni}. Out of which, four non-relativistic NMEs are extracted from wave-function expansion of p-wave leptons, and rest two relativistic NMEs  come out from the small components of Dirac spinors. These are

\begin{align}
	\begin{split}
	O(0^-) &: O_{\text{RA}}=\text{g}_{\rm A}(\bm{\sigma} \cdot \bm{p_e}),\,\,\, O_{\text{SA}}=\text{g}_{\rm A}(\bm{\sigma} \cdot\bm{ r}) \\
	O(1^-) &: O_{\text{RV}}=\text{g}_{\rm V}{\bm{p_e}},\,\,\, O_{\text{VA}}=\text{g}_{\rm A}(\bm{\sigma}\, \times \,\bm{ r}), \,\, O_{\text{VV}}=\text{g}_{\rm V}\bm{r} \\
	O(2^-) &: O_{\text{TA}}=\text{g}_{\rm A}[\bm{\sigma r}]_2\,,
	\end{split}
\end{align}
 where $O(0^-)$ is rank zero operator with $\Delta$J=0. The terms $O_{\text{RA}}$ and $O_{\text{SA}}$ are recoil-axial matrix element and scalar-axial matrix element, respectively. $O(1^-)$ is rank one operator with $\Delta$J=1. The terms $O_{\text{RV}}$, $O_{\text{VA}}$ and $O_{\text{VV}}$ are recoil-vector matrix element, vector-axial matrix element and vector-vector matrix element, respectively. $O(2^-)$ is rank two operator with $\Delta$J=2. The term $O_{\text{TA}}$ is tensor-axial matrix element. These all six operators changes parity during transition i.e. $\pi_i \pi_f=-1$. Further, the recoil-axial matrix element $\gamma_5$ gets enhanced over the impulse approximation with the aid of meson enhancement factor \cite{Towner} which is depicted by $\epsilon_{\text{MEC}}$. We have used the value $\epsilon_{\text{MEC}}$ = 2.01  of mesonic enhancement factor in these calculations for the rank zero nuclear matrix element $\gamma_5$ which corresponds to $\epsilon_{\text{MEC}}$ = 2.01$\pm$0.05 given in \cite{enhancement}.

We have obtained the quenching factor for $^{210-215}$Pb $\rightarrow$  $^{210-215}$Bi and $^{210-215}$Bi $\rightarrow$ $^{210-215}$Po transitions using the chi-square fitting method. We have compared theoretical and experimental average shape factor values for these transitions to get  the quenching factor. In our calculation for the  average shape factor, we have included the next-to-leading order terms \cite{mika2016}. First we have performed calculations without including the quenching factor by taking the bare values of the weak coupling constant, i.e., $g_A$=1.27 and $g_V$=1.00. Corresponding results are shown in 
Fig. 1(a); we can conclude that the theoretical and experimental values of the average shape factor are very far from each other.  Hence, we have used chi-square fitting method to obtain the quenching factor, 
it comes out to be 0.38 such that $g_A^{eff}=qg_A^{free}$= 0.4826 and $g_V^{eff}$ = 1.00. Using these values of weak coupling constants, we have plotted Fig. 1(b) for the same transitions. In this figure, the data points come close to the central line, which means  that theoretical values are approaching to the experimental ones. We have also calculated values of the average shape factor using another set of weak coupling constants taken from Ref. \cite{zhi}.  Corresponding results are shown in Fig. 1(c) for ($g_{\rm A}/g_{\rm A}^{\rm free}$, $g_{\rm V}/g_{\rm V}^{\rm free}$)=(0.38, 0.51). This figure shows the best fit for the average shape factor. In our further calculations, we have used these two sets of weak coupling constants i.e. set I: ($g_{\rm A}/g_{\rm A}^{\rm free}$, $g_{\rm V}/g_{\rm V}^{\rm free}$)=(0.38, 1.00) and set II: ($g_{\rm A}/g_{\rm A}^{\rm free}$, $g_{\rm V}/g_{\rm V}^{\rm free}$) = (0.38, 0.51).

\section{Results and Discussion}

\begin{table*}
\caption{Comparison between theoretical and experimental \cite{nndc} $\log ft$ values for Pb $\rightarrow$ Bi transitions. The calculations are carried out through two sets of quenching factor in the weak coupling constants $g_{\rm V}$ and $g_{\rm A}$,  for set I: ($g_{\rm A}/g_{\rm A}^{\rm free}$,$g_{\rm V}/g_{\rm V}^{\rm free}$)=(0.38,1.00) and set II: ($g_{\rm A}/g_{\rm A}^{\rm free}$,$g_{\rm V}/g_{\rm V}^{\rm free}$)=(0.38,0.51). The quenching factors of set I have been calculated from this work and set II have been taken from \cite{zhi}.}\label{tab:logft1} 
\begin{ruledtabular}
\begin{tabular}{lccccccccccc}
\multicolumn{2}{c}{~~~~~Transition} & Decay mode&Energy (keV) &\multicolumn{3}{c}{~~~~~$\log ft$} & \multicolumn{3}{c}{~~~~~$ {[\overline{C(w_{e})}]}^{1/2}$}  \T\B \\
\cline{1-2}
\cline{5-7}
\cline{8-10}

 Initial ($J_i^{\pi}$)& Final ($J_f^{\pi}$)  & & & Expt. & Set I &Set II & Expt. & Set I & Set II     \T\B\\\hline\T\B
			
 $^{210}$Pb$(0^+)$ & $^{210}$Bi$(1^-_1)$ & 1st FNU & 0.0&7.9(1)	&6.719 &8.517 & 3.436 & 13.380 & 1.689 \\ \T\B
  & $^{210}$Bi$(0^-_1)$ & 1st FNU & 46.539(1)&5.4(1) & 5.469 & 5.469 & 61.102 & 56.422 & 56.422 \\ \T\B
 $^{211}$Pb$(9/2^+)$ & $^{211}$Bi$(9/2_1^-)$ &1st FNU	& 0.0& 5.990(8) & 6.109 & 6.106 & 30.978 & 27.024 & 27.104\\ \T\B
  & $^{211}$Bi$(7/2^-_1)$ & 1st FNU	& 404.866(9)&7.19(3)	& 6.239 & 7.183 & 7.781 & 23.262 & 7.847 \\ \T\B
  & $^{211}$Bi$(9/2_2^-)$ & 1st FNU& 831.960(12)&5.7330(18) & 5.779 & 5.771 & 41.644  & 39.480 & 39.881 \\ \T\B
  & $^{211}$Bi$(9/2_3^-)$ & 1st FNU	& 1109.485(23)&5.58(4)	& 5.507 & 5.513 & 49.665 & 54.049 &53.646 \\ \T\B
$^{212}$Pb$(0^+)$ & $^{212}$Bi$(1_1^{(-)})$&1st FNU &0.0	& 6.73(4)	& 7.486 &7.288 & 13.215 & 5.534 & 6.952\\ \T\B
  & $^{212}$Bi$(2^{(-)}_1)$&1st FU 	& 115.183(5)	        & - & 10.143 & 10.140 & - & 0.260 &0.261 \\ \T\B
  & $^{212}$Bi$(0^{(-)}_1)$& 1st FNU & 238.632(2)&5.179	(10)& 5.208 & 5.208 & 78.805 & 76.179 & 76.179\\ \T\B
  & $^{212}$Bi$(1_2^{(-)})$	& 1st FNU& 415.272(11)& 5.342(17) & 4.551 & 5.156 & 65.321 & 162.418 & 80.915\\ \T\B
$^{213}$Pb$((9/2^+))$ & $^{213}$Bi$(9/2^-_1)$& 1st FNU&0.0	& 6.5	& 6.447 & 6.459 & 17.221  &18.297 & 18.057\\ \T\B
  & $^{213}$Bi$((7/2^-_1))$ & 1st FNU	& 257.63(7)&7.7	& 6.719 &7.887  &4.326  & 13.386  &3.489  \\ \T\B
  & $^{213}$Bi$((5/2^-_1))$ &1st FU & 592.72(8)	& 7.5	&13.355& 13.698 &5.446  &0.006&0.004\\ \T\B
  & $^{213}$Bi$((7/2^-_2))$ & 1st FNU & 592.72(8)	& 7.5	&8.184& 8.018 &5.446  &2.477&2.998\\ \T\B
  & $^{213}$Bi$((9/2^-_2))$ & 1st FNU &977.71(8)	& 5.6	& 6.128 & 6.107 & 48.535  & 26.423 & 27.065\\ \T\B
$^{214}$Pb$(0^+)$ & $^{214}$Bi$(1^-_1)$& 1st FNU&0.0	& 6.26(4)& 7.873 	& 7.144 & 22.701  &3.545 &8.205\\ \T\B
  & $^{214}$Bi$(2^-_1)$ & 1st FU & 53.2260(15)&	- & 9.446 & 9.445 & -  & 0.580 & 0.580\\ \T\B
  & $^{214}$Bi$((2^-_2))$	& 1st FU& 62.68(5)& -	& 8.087& 8.098& - &2.772&2.756 \\ \T\B 
  & $^{214}$Bi$((3^-_1))$	& 3rd FNU& 62.68(5)& -	& 15.860 & 14.949 & - &53.652 &153.140 \\ \T\B
  & $^{214}$Bi$((2^-_3))$	& 1st FU& 258.869(24) & 8.04(12) & 8.665	& 8.688 & 2.924& 1.424 &1.387 \\ \T\B
  & $^{214}$Bi$(1^-_2)$& 1st FNU	&295.2236(19)& 5.250(24) 	& 4.494& 5.014 &  72.620& 173.437& 95.335\\ \T\B
  & $^{214}$Bi$(0^-_1)$& 1st FNU& 351.9323(21)	& 5.07(3) 	& 5.080& 5.080 & 89.342 & 88.348& 88.348\\ \T\B
  & $^{214}$Bi$(1^-_3)$& 1st FNU	& 351.9323(21)& 5.07(3) 	& 5.074& 6.187 & 89.342 & 88.880& 24.689\\ \T\B
  & $^{214}$Bi$((2^-_4))$	& 1st FU&377.03(4)& -	& 10.211&10.197  & - &0.240& 0.244\\ \T\B
  & $^{214}$Bi$((1^-_4))$	& 1st FNU&533.672(14)& 6.23(4)	& 7.374& 7.881 &23.499 & 6.304 & 3.511\\ \T\B
  & $^{214}$Bi$(1^+_1)$	& Allowed & 838.994(22)& 4.43(9) 	& 4.238 & 4.238 &  0.483 & 0.597& 0.597 \\ \T\B
$^{215}$Pb$((9/2^+))$ & $^{215}$Bi$((9/2^-_1))$&	1st FNU& 0.0&$\geq$6.1	& 6.713 &6.741  & $\leq$27.293 & 13.472&  13.045\\ \T\B
  & $^{215}$Bi$((7/2^-_1))$&	1st FNU &183.5(3)	& $>$6.6	&7.224& 8.486 & $<$15.348 &7.480 & 1.751
\end{tabular}
\end{ruledtabular}
\end{table*}

In the present work, large-scale shell-model calculations for the allowed and forbidden beta-decay transitions have been carried out in the north-east region of doubly-magic nucleus $^{208}$Pb and compared these results with the recently available experimental data  \cite{nndc}. In this work, for the first time, we present theoretical $\log ft$ values, average shape factors, and half-lives for $^{210-215}$Pb $\rightarrow$ $^{210-215}$Bi and $^{210-215}$Bi $\rightarrow$ $^{210-215}$Po transitions.  We have also calculated the quenching factor (as discussed in  the above section) and theoretical Q-value for these nuclei, and we have compared the shell-model results with the experimental data.
In the present work we have performed shell-model calculations by taking $^{208}$Pb as a core using KHPE effective interaction \cite{war91} given by Kuo Herling for $^{208}$Pb (Z=82, N=126). Model space for this interaction is $0h_{9/2}, 1f_{7/2}, 1f_{5/2}, 2p_{3/2}, 2p_{1/2}, 0i_{13/2}$ for protons (82 $ < $ Z $<$ 126) and $0i_{11/2}, 1g_{9/2}, 1g_{7/2}, 2d_{5/2}, $ $2d_{3/2}, 3s_{1/2}, 0j_{15/2}$ for neutrons (126 $ < $ N $<$ 184).

In Table \ref{tab:logft1}, $\log ft$ values and average shape factor have been calculated from the ground states of $^{210-215}$Pb to the ground, and several excited states of $^{210-215}$Bi using both sets of the weak coupling constants.  At some places, double bracket represents those states which are not confirmed experimentally. These results have been computed using experimental Q- values taken from \cite{nndc}. On comparing experimental and theoretical results in Table \ref{tab:logft1}, we notice that both sets of weak coupling constants give promising results for the $\log ft$ values. As we can observe that $\log ft$ values for  the transition from $^{210}$Pb($0^+$) to $^{210}$Bi($0^-_1$) at energy 46.539(1) keV is 5.469 for both of the sets which agrees quite well the experimental value i.e. 5.4(1). In the case of $^{211}$Pb(9/2$^+$) to $^{211}$Bi(9/2$_2^-$) transition at energy 831.960(12) keV, the theoretical value of $\log ft$ is 5.779 for the set I and 5.771 for the set II, which is near to the experimental value, i.e., 5.7330(18). Shell-model results for $^{211}$Pb to $^{211}$Bi transitions have also been calculated by Warburton \cite{enhancement} with KHPE interaction using old data. Further, for  the transition  from $^{213}$Pb((9/2$^+$)) to $^{213}$Bi(9/2$^-_1$) theoretical $\log ft$ value for the set I is 6.447 and for the set II is 6.459, while the experimental value is  6.5. In the case of  $^{215}$Pb((9/2$^+$)) to $^{215}$Bi((9/2$^-_1$)) transition, experimentally, the $\log ft$ value should be greater than or equal to 6.1, which is confirmed by shell-model results as $\log ft$ value for the set I is 6.713 and for the set II is 6.741. Also, in  the  $^{215}$Pb((9/2$^+$)) to $^{215}$Bi((7/2$^-_1$)) transition at energy 183.5(3) keV, this $\log ft$ value should be greater than 6.6, which can be verified via theoretical results, i.e., for the set I, it is 7.224, and for  the set II it is 8.486. There are several such transitions for which theoretical $\log ft$ values match very well with the experimental values, which shows the authenticity of shell-model calculations. We have also calculated  $\log ft$ results corresponding to the transitions for which experimental data is unavailable. For instance, $\log ft$ value for $^{212}$Pb(0$^+$) to $^{212}$Bi(2$^{(-)}_1$) transition at energy 115.183(5) keV is 10.143 for  the set I and 10.140 for  the set II. In case of $^{214}$Pb(0$^+$) to $^{214}$Bi($2^-_1$) transition at energy 53.2260(15) keV, this value comes out to be 9.446 for  the set I and 9.445 for  the set II while for $^{214}$Pb(0$^+$) to $^{214}$Bi(($2^-_4$)) transition at energy 377.03(4) keV, this value comes out to be 10.211 for the set I and 10.197 for the set II. Furthermore, in case of $^{213}$Pb((9/2$^+$)) $\rightarrow$ $^{213}$Bi((7/2$^-_2$),(5/2$^-_1$)) at energy 592.72(8) keV, experiments  do not confirm one suitable spin state thus predict two possible spin states. Therefore, we have performed shell-model calculations for both possible spin states $7/2^-_2$ and $5/2^-_1$. It is inferred from these results that the  spin and parity of the state at 592.72(8) keV is $7/2^-_2$ as its $\log ft$ value is 8.184 for the set I and 8.018 for  the set II, which is close to the experimental value, i.e., 7.5. However, for the  $^{214}$Pb(0$^+$) to $^{214}$Bi(0$^-_1 $,1$^-_3$) transition at energy 351.9323(21) keV, our calculated shell-model results for the $\log ft$ values  give good results for both possible spins of the state. Thus it is difficult to distinguish  the spin-parity of these two states. There are some transitions in this table for which shell-model results overestimates or underestimates the experimental $\log ft$ values. For instance, for $^{210}$Pb($0^+$) to $^{210}$Bi($1^-_1$) transition, the experimental $\log ft$ value is 7.9(1), while for the set I, it is 6.719, and for  the  set II it is 8.517. Similarly, for $^{212}$Pb(0$^+$) to $^{212}$Bi(1$^{(-)}_1$), the experimental value is 6.73(4), while it is 7.486 for the set I and it is 7.288 for  the set II. 

\begin{table*}
\centering
\caption{The same as in Table \ref{tab:logft1} for the Bi$\to\,$Po transitions.} \label{tab:logft2}
\begin{ruledtabular}
\begin{tabular}{lccccccccccc}
\multicolumn{2}{c}{~~~~~Transition} &Decay mode & Energy (keV)& \multicolumn{3}{c}{~~~~~$\log ft$} & \multicolumn{3}{c}{~~~~~$ {[\overline{C(w_{e})}]}^{1/2}$}  \\
\cline{1-2}
\cline{5-7}
\cline{8-10}
 Initial ($J_i^{\pi}$)& Final ($J_f^{\pi}$)& & & Expt. & Set I&Set II & Expt. & Set I & Set II     \T\B\\\hline\T\B
			
$^{210}$Bi$(1^-)$ & $^{210}$Po$(0^+_1)$ &	1st FNU&0.0	& 8.0	&6.690& 8.114&3.062 & 13.842&2.686  \\ \T\B
$^{211}$Bi$(9/2^-)$		& $^{211}$Po$(9/2^+_1)$ &	1st FNU&0.0& 5.99(2)	& 6.143 & 6.140 & 30.978  & 25.982& 26.053	\\ \T\B
$^{212}$Bi$(1^{(-)})$	& $^{212}$Po$(0^+_1)$&	1st FNU&0.0	& 7.2664(16)	& 7.330&8.951  & 7.126 & 6.621& 1.024\\ \T\B
			& $^{212}$Po$(2^+_1)$ 	&	1st FNU	&727.330(9) & 7.720(11)	& 7.017& 7.609 & 4.227 & 9.498&4.803 \\ \T\B
                       & $^{212}$Po$(2^+_2)$ &	1st FNU&1512.70(8)	& 7.093(13)	& 6.554& 6.645 & 8.701 & 16.179& 14.581\\ \T\B
                       & $^{212}$Po$(1^+_1)$&	1st FNU&1620.738(10)	& 6.748(11)	& 6.079& 6.384 & 12.944& 27.973& 19.679\\ \T\B
                       & $^{212}$Po$(2^+_3)$ &	1st FNU&1679.450(14)	& 7.51(6)	& 5.850& 6.447 & 5.383  & 36.411& 18.295\\ \T\B
	    	       & $^{212}$Po$(0^+_2)$ &	1st FNU&1800.9(2)		& 8.05(9)	& 8.103& 7.780 & 2.891  & 2.721&3.945 \\ \T\B
                      & $^{212}$Po$(2^+_4)$ &	1st FNU&1805.96(10)	& 6.695(21)	& 6.546& 6.926 & 13.758 & 16.327& 10.543\\ \T\B
$^{213}$Bi$(9/2^-)$	& $^{213}$Po$(9/2^+_1)$ &	1st FNU&	0.0& 6.31(1)& 6.367 	& 6.358 & 21.235  & 20.068& 20.287 \\ \T\B		   				& $^{213}$Po$((11/2^+_1))$	&	1st FNU&292.805(8)	& 8.45(10)	& 5.386&5.869  & 1.910 & 62.110& 35.590\\ \T\B
			& $^{213}$Po$((7/2^+_1))$&	1st FNU	&440.446(9)	& 6.08(1)	& 9.381& 9.302 & 28.252  & 0.624&0.684 \\ \T\B
		        & $^{213}$Po$((5/2^+_1))$	&	1st FU&600.87 (17)	& 10.03(9)	& 13.513& 13.462 & 0.448 & 0.005& 0.006\\ \T\B
			& $^{213}$Po$((13/2^+_1))$&	1st FU&867.98(3)	& 8.64(5)	& 10.457& 10.458 & 1.589 & 0.181& 0.181 \\ \T\B			
			& $^{213}$Po$((9/2^+_2))$ &	1st FNU&1003.605(22)	& 7.49(3)& 6.096	& 6.091 & 5.452 & 27.435&27.586 \\ \T\B			
		& $^{213}$Po$((9/2^+_3))$&	1st FNU& 1045.65(9)& 7.85(7)	& 7.196 & 7.211 & 3.640 &7.727&7.592 \\ \T\B
		& $^{213}$Po$((11/2^+_2))$&	1st FNU& 1045.65(9)& 7.85(7)	& 8.293 & 8.591 & 3.640 &2.184&1.551 \\ \T\B
$^{214}$Bi$(1^-)$	& $^{214}$Po$(0^+_1)$ &	1st FNU	&0.0	& 7.872(11)&7.314	&8.311 &3.586 &6.746 &2.142\\ \T\B
		        & $^{214}$Po$(2^+_1)$ 	&	1st FNU&609.318(5)	& 9.06(7)	& 7.649&  8.318& 0.904 & 4.590& 2.124\\ \T\B
			& $^{214}$Po$(3^-_1)$ &	2nd FNU&1274.765(9)	& 9.5(3)	& 11.161&11.542 & 210.292 & 31.069& 20.037\\ \T\B
			& $^{214}$Po$(2^+_2)$ &	1st FNU&1377.681(7)	& 7.374(11)	& 6.695 & 6.978& 6.296  & 13.754& 9.934\\ \T\B
			& $^{214}$Po$(0^+_2)$ &	1st FNU&1415.498(8)	& 8.25(3)	& 7.380& 7.644 & 2.296 & 6.250& 4.616\\ \T\B
			& $^{214}$Po$(2^+_3)$ &	1st FNU&1661.282(14)	& 8.21(4)	& 6.302 & 6.929 & 2.405 & 21.624&10.510 \\ \T\B
		        & $^{214}$Po$(2^+_4)$ &	1st FNU&1729.613(7)		& 6.654(12) & 6.713 &  6.240& 14.423 & 13.470 &23.241\\ \T\B
		        & $^{214}$Po$(2^+_5)$ &	1st FNU	&1847.446(9)	& 6.859(13)& 6.509	&7.170 & 11.391 & 17.034& 7.961\\ \T\B
			& $^{214}$Po$((2^+_6))$ &	1st FNU&2010.831(13) & 7.422(15) & 6.381	&6.730 & 5.957 &19.749&13.212\\ \T\B
$^{215}$Bi$((9/2^-))$	& $^{215}$Po$(9/2^+_1)$	&	1st FNU&0.0& $>$6.9&6.653	& 6.642& $<$10.866 & 14.442 & 14.619\\ \T\B
			& $^{215}$Po$(7/2^+_1)$ &	1st FNU&271.11(10)	& $>$8.2 & 10.020	& 8.737 & $<$2.432 & 0.299 & 1.312\\ \T\B
	               & $^{215}$Po$((11/2^+_1))$	&	1st FNU&293.53(10)	& 6.0(1)&5.349	& 5.813& 30.623 & 64.770 &38.001  \\ \T\B
	               & $^{215}$Po$(5/2^+_1)$	&	1st FU&401.6(10)	& 7.7	& 13.796& 13.714 & 4.326 & 0.004& 0.004\\ \T\B
	        & $^{215}$Po$(7/2^+_2)$ &	1st FNU&517.53(17)& 7.8	& 9.041& 8.965&3.855 & 0.924& 1.008\\ \T\B
	        & $^{215}$Po$(9/2^+_2)$ &	1st FNU&517.53(17)& 7.8	& 6.619& 6.608&3.855 & 15.024& 15.20\\ \T\B
	      & $^{215}$Po$((11/2^+_2))$ &1st FNU&609.0(5)	& 7.4	&9.058&10.090 &6.110 &0.905&0.276 \\ \T\B
	      & $^{215}$Po$((13/2^+_1))$ & 1st FU	&609.0(5)& 7.4	&9.931&9.933 &6.110 &0.332&0.331
			    			    			    			
		\end{tabular}
	\end{ruledtabular}
\end{table*}

Table \ref{tab:logft2} shows the $\log ft$ and average shape factor values for the transitions from the ground states of $^{210-215}$Bi to the ground and different excited states of $^{210-215}$Po, calculated using both sets of the weak coupling constants. Decay mode and experimental data are also included in this table. The theoretical results are in quite good agreement with the experimental data. For instance, the $\log ft$ value for $^{210}$Bi$(1^-)$ to $^{210}$Po$(0^+_1)$ transition is 8.114 for the set II, which is close to the experimental value i.e. 8.0. In the case of $^{211}$Bi$(9/2^-)$ to $^{211}$Po$(9/2^+_1)$ transition, the experimental $\log ft$ value is 5.99(2), which is close to the shell-model results; i.e., for the set I, it is 6.143, and for the set II it is 6.140. Also for $^{212}$Bi$(1^{(-)})$ to $^{212}$Po$(2^+_4)$ transition at energy 1805.96(10) keV, the theoretical $\log ft$ value for the set I is 6.546 and for the set II it is 6.926, which matches with the experimental value i.e. 6.695(21). For $^{213}$Bi$(9/2^-)$ to $^{213}$Po(9/2$^+_1$) transition, this value is 6.367 for the set I and 6.358 for the set II which is very near to the experimental value, i.e., 6.31(1). On moving toward the transition from $^{214}$Bi$(1^-)$	to $^{214}$Po$(0^+_1)$, the experimental $\log ft$ value is 7.872(11), which matches with shell-model results, i.e., for the set I, it is 7.314, and for the set II it is 8.311. Further, in $^{215}$Bi$((9/2^-))$	to $^{215}$Po$((11/2^+_1))$ transition at energy 293.53(10) keV, the experimental and shell-model results for the set II are in good agreement with each other, i.e., experimentally $\log ft$ value is 6.0(1) while for the set I, it is 5.349 and for the set II it is 5.813. There are several transitions for which our computations overestimate or underestimate the $\log ft$ value. For instance, there are small discrepancies in the results of $^{213}$Bi decay. For the transition from $^{213}$Bi$(9/2^-)$	to $^{213}$Po$((5/2^+_1))$ at energy 600.87(17) keV, the experimental $\log ft$ value is 10.03(9) whereas the theoretical value for the set I is 13.513 and for the set II is 13.462. Also, for $^{215}$Bi((9/2$^-$)) to $^{215}$Po(5/2$^+_1$) transition at energy 401.6(10) keV, the experimental $\log ft$ value is 7.7 while the theoretical value for the set I is 13.796, and for the set II is 13.714, which are approximately twice the experimental value. It can be because of several reasons. Firstly, the 9/2$^-$ state of $^{215}$Bi is not yet confirmed, and secondly, there are several approximations (such as impulse approximation) assumed while calculating $\log ft$ and half-lives in the shell-model calculations. Further, there are several transitions where unique assignments of spin-parity are not possible. We have calculated shell-model results for all possible spins and parities of the states. However, it is difficult to make unique assignments for these states because their results are in close proximity to each other. For instance, in the case of $^{213}$Bi$(9/2^-)$	to $^{213}$Po$((9/2^+_3),(11/2^+_2))$ at energy 1045.65(9) keV, the experimental $\log ft$ value is 7.85(7), and the shell-model results for both $9/2^+_3$ and $11/2^+_2$ in $^{213}$Po give values close to the experimental value. Thus, we can not make a unique assignment of spin-parity for the state of $^{213}$Po at 1045.65(9) keV.    

\begin{table*}
\caption{Comparison between the shell-model Q-values and experimental \cite{nndc} Q-values.}\label{tab:Qval}\T\B
\begin{ruledtabular}
\begin{tabular}{lccccccccccc}
\multicolumn{2}{c}{~~~~~Transition} &\multicolumn{2}{c}{~~~~~E (SM) (MeV)} &\multicolumn{2}{c}{~~~~~Q-value (MeV)}  \T\B \\
\cline{1-2}
\cline{3-4}
\cline{5-6}

 Initial& Final  & Initial& Final & Expt.  & Theoretical     \T\B\\\hline\T\B
			
 $^{210}$Pb$(0^+)$ & $^{210}$Bi$(1^-)$ & -9.091 &-8.403	& 0.0635(5)  &0.092  \\ \T\B
 $^{211}$Pb$(9/2^+)$ & $^{211}$Bi$(9/2^-)$ &-12.936	& -13.512 &1.367(6)   &1.356 \\ \T\B
 $^{212}$Pb$(0^+)$ & $^{212}$Bi$(1^{(-)})$ &-18.034	& -17.882 &0.5691(18)  &0.628 \\ \T\B
 $^{213}$Pb$((9/2^+))$ & $^{213}$Bi$(9/2^-)$& -21.762 & -23.044  & 2.030(8) &2.062 \\ \T\B
 $^{214}$Pb$(0^+)$ & $^{214}$Bi$(1^-)$ & -26.788 & -27.149 &1.018(11)  &1.141 \\ \T\B
  $^{215}$Pb$((9/2^+))$ & $^{215}$Bi$((9/2^-))$ & -30.370 & -32.355 &2.770(10) & 2.765 \\ \T\B
 
 $^{210}$Bi$(1^-)$& $^{210}$Po$(0^+)$ &-8.403	& -8.762	&1.1622(8)   &1.139\\ \T\B
$^{211}$Bi$(9/2^-)$&$^{211}$Po$((9/2^+))$ & -13.512	& -13.309	&0.574(5)   & 0.577\\ \T\B
 $^{212}$Bi$(1^{(-)})$& $^{212}$Po$(0^+)$ &-17.882	& -19.212	&2.2515(17)   &2.110\\ \T\B
 $^{213}$Bi$(9/2^-)$& $^{213}$Po$(9/2^+)$ & -23.044 &-23.598  & 1.422(5) &  1.334\\ \T\B	
 $^{214}$Bi$(1^-)$	& $^{214}$Po$(0^+)$ & -27.149 &-29.440   & 3.269(11)&3.071\\ \T\B
 $^{215}$Bi$((9/2^-))$	& $^{215}$Po$(9/2^+)$ &-32.355 &-33.661 &2.189(15) &2.086 

\end{tabular}
\end{ruledtabular}
\end{table*}

As Q-values play very important roles in the calculation of beta-decay properties, we have also calculated theoretical Q-values for the concerned nuclei which are listed in Table \ref{tab:Qval}. In this paper, we have used experimental Q-values for the calculation of shell-model results. Further, for better comparison, we have also used theoretical Q-values  to calculate $\log ft$ and half-lives of the transitions included in this paper. Column I and II show initial and final ground states of the transitions of the concerned nuclei, and column III and IV shows shell-model binding energies of these nuclei. Column V shows experimental Q-values and column VI show calculated shell-model Q-values. It can be concluded that these shell-model Q values are in close proximity with the experimental ones. For instance, the theoretical Q-value for $^{215}$Pb to $^{215}$Bi transition is 2.765 MeV while the experimental value is 2.770(10) MeV. These two values are very close to each other.

\begin{figure}
  \includegraphics[width=95mm]{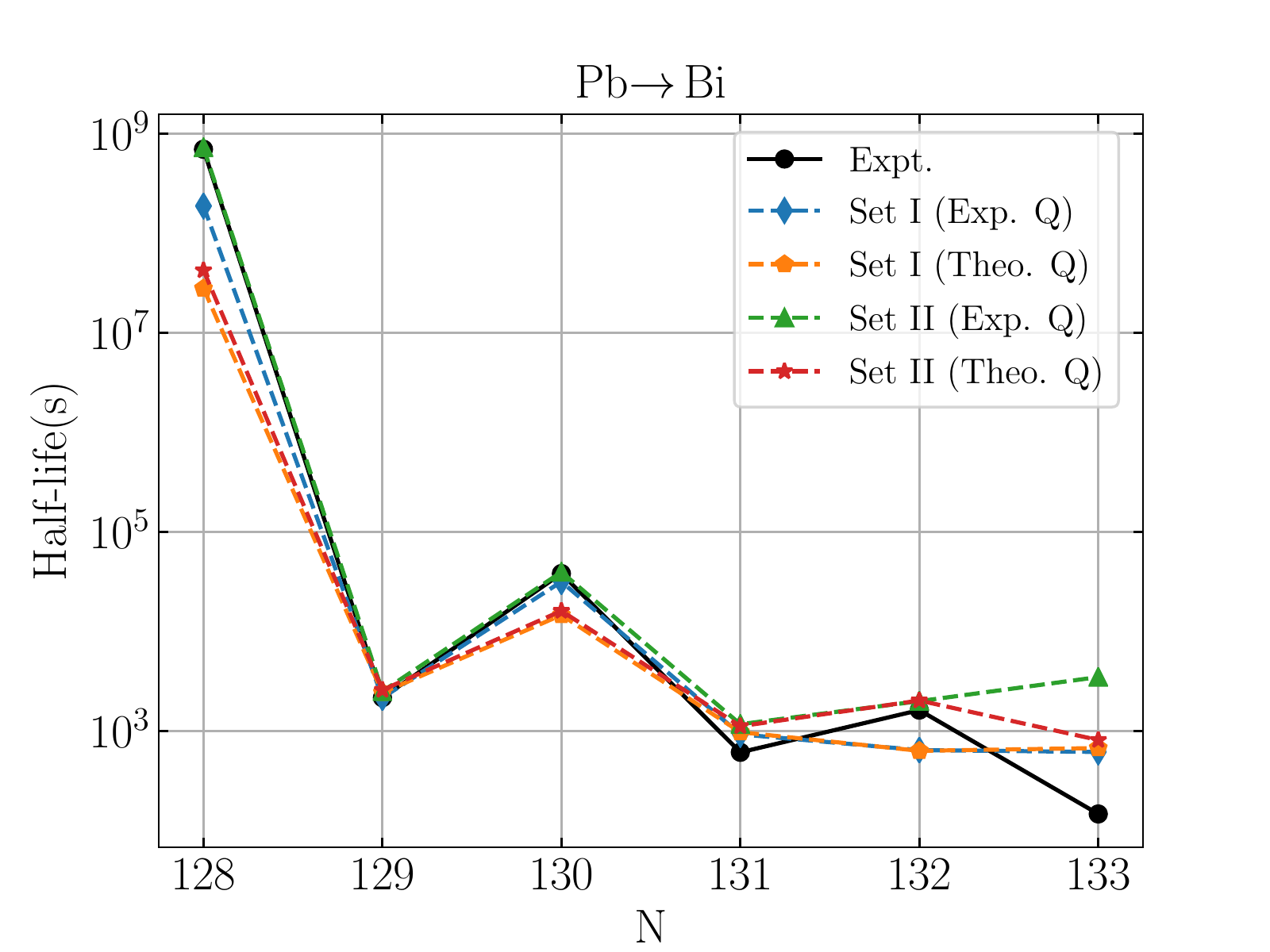}
\caption{\label{fig:h1}Comparison of calculated and experimental \cite{nndc} half-lives for allowed and first-forbidden transitions for Pb $\rightarrow$ Bi with experimental and shell-model Q-values.  For set I : ($g_A/g_A^{free}$, $g_V/g_V^{free}$)=(0.38, 1.00) and set II :  ($g_A/g_A^{free}$, $g_V/g_V^{free}$)=(0.38, 0.51) have been taken.}
\end{figure}

\begin{figure}
  \includegraphics[width=95mm]{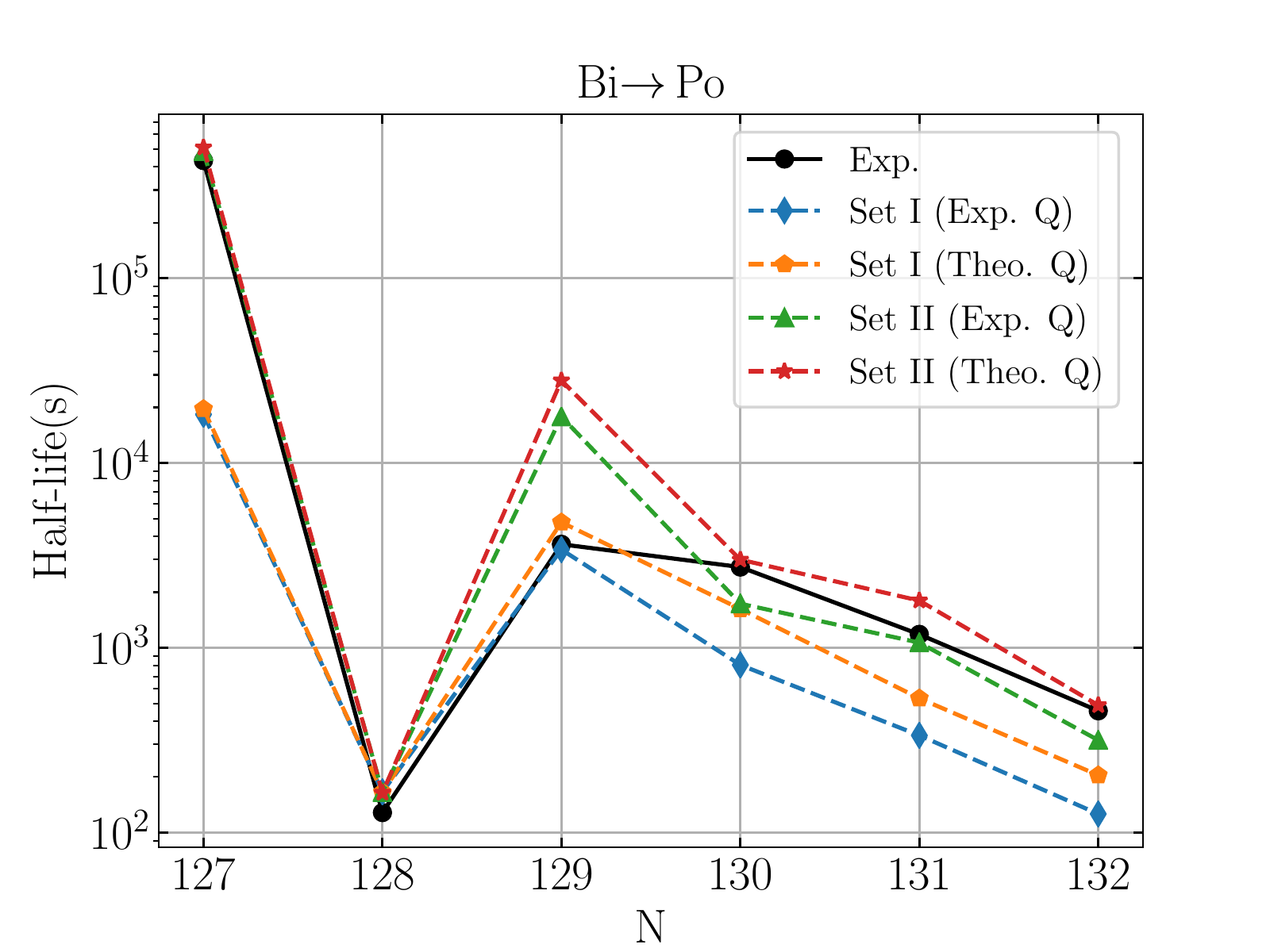}
\caption{\label{fig:h2}Comparison of calculated and experimental \cite{nndc} half-lives for allowed and first-forbidden transitions for Bi $\rightarrow$ Po with experimental and shell-model Q-values.  For set I : ($g_A/g_A^{free}$, $g_V/g_V^{free}$)=(0.38, 1.00) and set II :  ($g_A/g_A^{free}$, $g_V/g_V^{free}$)=(0.38, 0.51) have been taken.}
\end{figure}

\begin{table*}
	\centering
	\caption{Comparison between theoretical and experimental \cite{nndc} half-lives values for Pb $\rightarrow$ Bi transitions for experimental and theoretical Q-values. The calculations are carried out through two sets of quenching factors in the weak coupling constants $g_{\rm V}$ and $g_{\rm A}$. The values of quenching factor for set I: ($g_{\rm A}/g_{\rm A}^{\rm free}$, $g_{\rm V}/g_{\rm V}^{\rm free}$)=(0.38, 1.00) and set II: ($g_{\rm A}/g_{\rm A}^{\rm free}$, $g_{\rm V}/g_{\rm V}^{\rm free}$)=(0.38, 0.51). The quenching factors of set I have been calculated from this work and set II have been taken from \cite{zhi}.}\label{tab:half1} 
	\begin{ruledtabular}
\begin{tabular}{lccccccccccc}
\multicolumn{2}{c}{~~~~~Transition}   & \multicolumn{5}{c}{Half-life}   \\
			\cline{1-2}
			\cline{3-7}

Initial& Final   & Expt. & Set I (Exp. Q) & Set I (Theo. Q)&Set II (Exp. Q) & Set II (Theo. Q)      \T\B\\\hline\T\B

 $^{210}$Pb$(0^+)$ & $^{210}$Bi	& 22.20(22) y &  5.974 y & 0.886 y& 23.087 y & 1.350 y \\ \T\B
$^{211}$Pb$(9/2^+)$ & $^{211}$Bi	 & 36.1(2) min& 36.052 min & 40.534 min& 41.743 min& 43.157 min \\ \T\B
$^{212}$Pb$(0^+)$ & $^{212}$Bi 	 & 10.622(7) h  & 8.804 h&4.069 h &11.004 h & 4.476 h \\ \T\B
$^{213}$Pb$((9/2^+))$ & $^{213}$Bi	 & 10.2(3) min & 15.452 min & 16.246 min& 19.542 min& 18.661 min\\ \T\B
$^{214}$Pb$(0^+)$ & $^{214}$Bi  & 27.06(7) min & 10.7329 min &10.558 min& 33.306 min &33.792 min\\ \T\B
 $^{215}$Pb$((9/2^+))$ & $^{215}$Bi  & 147 s &617.223 s& 675.112 s& 3486.387 s & 813.294 s 

		\end{tabular}
	\end{ruledtabular}
\end{table*}

\begin{table*}
	\centering
	\caption{The same as in Table \ref{tab:half1} for the Bi$\to\,$Po transitions.} \label{tab:half2} 
	\begin{ruledtabular}
\begin{tabular}{lccccccccccc}
\multicolumn{2}{c}{~~~~~Transition}   & \multicolumn{5}{c}{Half-life}   \\
			\cline{1-2}
			\cline{3-7}

Initial& Final  & Expt. & Set I (Exp. Q) & Set I (Theo. Q)&Set II (Exp. Q) & Set II (Theo. Q)     \T\B\\\hline\T\B

$^{210}$Bi$(1^-)$ & $^{210}$Po & 5.012(5) d& 0.212 d &0.227 d& 5.631 d& 5.894 d\\ \T\B
$^{211}$Bi$(9/2^-)$& $^{211}$Po & 2.14 (2) min & 2.769 min &2.750 min & 2.754 min&2.736 min\\ \T\B
$^{212}$Bi$(1^{(-)})$	& $^{212}$Po & 60.55(6) min & 56.694 min&79.610 min & 297.021 min&466.002 min\\ \T\B
$^{213}$Bi$(9/2^-)$& $^{213}$Po  & 45.59(6) min &  13.461 min&27.040 min & 28.880 min &49.984 min\\ \T\B	
$^{214}$Bi$(1^-)$	& $^{214}$Po 	& 19.71(2) min  &5.598 min& 8.885 min&17.775 min&30.015 min\\ \T\B
$^{215}$Bi$((9/2^-))$	& $^{215}$Po  & 7.6(2) min& 2.103 min& 3.409 min& 5.275 min&8.152 min
\end{tabular}
	\end{ruledtabular}
\end{table*}

In Table \ref{tab:half1}, a comparison between theoretical and experimental half-lives has been given for Pb to Bi transitions using both the sets of the coupling constants. We have calculated half-lives using experimental Q-values. As we know, beta-decay half-lives are very sensitive to the Q-values, we have calculated half-lives using theoretical Q-values also. These half-lives have been calculated using the property of transition probability that it is additive in nature. It can be concluded from this table that our calculations of half-lives match well with the experimental half-lives in most of the cases. For instance, experimental half-life value for $^{211}$Pb to $^{211}$Bi transition is 36.1(2) min while theoretical value for the set I is 36.052 min with experimental Q-value. In case of the transition from $^{210}$Pb$(0^+)$ to $^{210}$Bi, experimental half-life does not match well with the result of the set I but matches well with the result of the set II. Minor discrepancies in the results can be explained as there are several transitions for which unique spin-parity assignments are not possible. Therefore, we have to exclude these transitions in our calculations, leading to some deviation in the results.

Table \ref{tab:half2} compares theoretical and experimental half-lives for Bi to Po transitions using both sets of the weak coupling constants. Calculations for half-lives using both theoretical and experimental Q-values have been done. The theoretical results for the half-lives agree pretty well with the experimental half-lives in most of the cases. For instance, experimental half-life value for $^{210}$Bi to $^{210}$Po transition is 5.012(5) d while theoretical value for the set II is 5.631 d with the experimental Q-value and it comes out to be 5.894 d with the theoretical Q-value.

For better comparison of half-lives, we have plotted calculated and experimental half-lives with the use of both the experimental and shell-model Q-values in Fig. \ref{fig:h1} on the log scale for Pb to Bi transitions. This plot shows half-lives for both the sets of the weak coupling constants using both experimental and theoretical Q-values. It can be concluded from Fig. \ref{fig:h1}  here that the set II with the experimental Q values matches extremely well with the experimental half-lives up to N=130. Further, at N=131, there is a good agreement between the experimental half-lives and those obtained for the set I with both the experimental and theoretical Q-values. There is a slight deviation at N=133. However, the set I using both the experimental and theoretical Q values approach the experimental value significantly better than others.

In Fig. \ref{fig:h2}, half-lives have been plotted for Bi to Pb transitions for both sets of the weak coupling constants using both Q-values. For N=127 and 128, results of the set II obtained with experimental and theoretical Q- values agree very well with the experimental results. However, the set I using experimental Q-values is more favorable than others at N=129, while there is a good agreement between the experimental values and set II values for both theoretical and experimental Q- values at N=130 to 132.

\begin{table*}
\caption{Error in calculated half-life values in comparison to experimental \cite{nndc} half-lives. Comparison is given for both sets of  the weak coupling constants using both experimental and shell-model Q-values.}\label{tab:err}\T\B
\begin{ruledtabular}
\begin{tabular}{lccccccccccc}
 &\multicolumn{2}{c}{~~~Using Q (Exp.)} &\multicolumn{2}{c}{~~~Using Q (Theo.)}  \T\B \\
\cline{2-3}
\cline{4-5}

   & Set I& Set II & Set I  & Set II     \T\B\\\hline\T\B
			
 $\bar{r}$ &-0.264  &0.191	&  -0.279 & 0.075 \\ \T\B
 $\sigma$ &0.48	& 0.423 & 0.576  & 0.487\\ \T\B
  $10^{\bar{r}}$&0.277	& 0.694 & 0.449 & 1.073\\ \T\B
 $10^\sigma$& 3.022 & 2.646  & 3.772 & 3.067

\end{tabular}
\end{ruledtabular}
\end{table*}

For a better and more detailed comparison of the shell-model and experimental half-lives, we have calculated the error in the theoretical half-lives compared to the experimental ones. Table \ref{tab:err} shows the deviation in calculated half-lives compared to the experimental half-lives on the log scale because the magnitude of half-lives listed in Tables \ref{tab:half1} and \ref{tab:half2} vary in a wide range. Therefore in this table, we have tried to show the mean deviation and fluctuation \cite{yoshida,moller03,mark16} from the experimental data on the log scale for both sets of the weak coupling constants, including both theoretical and experimental Q-values. Here, $r$ is the measure of deviation which is defined as,

\begin{equation}
	r=\log_{10}(T_{1/2}^{\rm{Calc}}/T_{1/2}^{\rm{exp}}),
\end{equation}
where $T_{1/2}^{\rm{Calc}}$ is the theoretically calculated half-life value and $T_{1/2}^{\rm{exp}}$ is the experimental half-life value. Its mean value is denoted by $\bar{r}$ and standard deviation is denoted by $\sigma$. These both quantities are defined as,

\begin{equation}
	\bar{r}=\frac{1}{n}\sum_{i=1}^n r_i,
\end{equation}

\begin{equation}
	\sigma=\Big[\frac{1}{n}\sum_{i=1}^n(r_i-\bar{r})^2\Big]^{1/2},
\end{equation}
 where $n$ is the number of transitions taken to calculate the half-lives and index $i$ goes from 1 to $n$. In these calculations, the value of $n$ is 12.
The mean error value and standard deviation should be equal to zero for minimal deviations because we have calculated it on the log scale. Our results are approaching zero for both experimental and theoretical Q-values, so we can conclude that our calculations show good agreement with the experimental data. Use of experimental Q-values is generally better than the case of theoretical Q-values. Use of the set II improves the agreement with the experimental data compared with the set I. We have also shown the mean and standard deviation in powers of 10 to eliminate the log scale.

\begin{table*}
\caption{ Comparison between theoretical and experimental \cite{iso} $\log ft$ values for $^{214}$Bi$^{\rm m}((8^-))$ to the different excited states in $^{214}$Po transitions. The calculations are carried out through two sets of quenching factor in the weak coupling constants $g_{\rm V}$ and $g_{\rm A}$. The values of quenching factor for set I: ($g_{\rm A}/g_{\rm A}^{\rm free}$, $g_{\rm V}/g_{\rm V}^{\rm free}$)=(0.38, 1.00) and set II: ($g_{\rm A}/g_{\rm A}^{\rm free}$, $g_{\rm V}/g_{\rm V}^{\rm free}$)=(0.38, 0.51). The quenching factors of set I have been calculated from this work and set II have been taken from \cite{zhi}.}\label{tab:iso1} 
\begin{ruledtabular}
\begin{tabular}{lccccccccccc}
\multicolumn{2}{c}{~~~~Final state} & Decay mode & Energy (keV)&\multicolumn{3}{c}{~~~~~$\log ft$} & \multicolumn{3}{c}{~~~~~$ {[\overline{C(w_{e})}]}^{1/2}$} \T\B \\
\cline{1-2}
\cline{5-7}
\cline{8-10}

  $J_f^{\pi}$ (Expt.) & $J_f^{\pi}$ (SM) & & & Expt. & Set I &Set II & Expt. & Set I & Set II     \T\B\\\hline\T\B
			
 $(8^+)$&$8^+_1$ & 1st FNU & 1584.4(4)&6.25(12)	&6.264 &6.255 & 22.964 & 22.587 & 22.834 \\ \T\B
   $(8^+,9^+)$&$8^+_2$& 1st FNU&1633.5(4) & 6.62(14) & 6.466 & 6.518 &14.999  & 17.917 & 16.867 \\ \T\B
 $(8^+,9^+)$&$9^+_1$&1st FNU	& 1633.5(4)&6.62(14) & 5.468 & 5.941 & 14.999 & 56.523 & 32.786\\ \T\B
  $(8^+)$ & $8^+_3$ & 1st FNU	& 1824.5(4)&6.25(12)	& 5.945 & 6.018 & 22.964 & 32.607 & 30.006 \\ \T\B
   $(6^+,7^+)$ &$6^+_1$ & 1st FU& 1843.0(4)&8.69(17) & 11.130 & 11.147 & 1.384  & 0.083 & 0.082 \\ \T\B
   $(6^+,7^+)$ &$7^+_1$ & 1st FNU&1843.0(4)	& 7.86(13)	& 9.643 & 8.726 & 3.598 & 0.462 &1.328 \\ \T\B
 $(8^+,9^+)$& $8^+_4$&1st FNU &1969.1(4)	& 6.93(13)	& 6.371 &6.757 &  10.497& 19.973 & 12.813\\ \T\B
 $(8^+,9^+)$&   $9^+_2$&1st FNU 	&1969.1(4)& 	 6.93(13) & 7.005&7.517 &  10.497& 9.631 &5.341 \\ \T\B
    $(8^+,9^+)$&$8^+_5$& 1st FNU &2059.5(4)& 6.04(13)	& 6.771 & 7.558 & 29.245 & 12.610 & 5.096\\ \T\B
  $(8^+,9^+)$&  $9^+_3$	& 1st FNU&2059.5(4)& 6.04(13) & 10.183 & 10.148 & 29.245 & 0.248 & 0.258\\ \T\B
$(9)$& $9^+_4$& 1st FNU	& 2159.0(4)&7.80(16)	& 7.743 & 7.895 &  3.855 &4.118 & 3.454\\ \T\B
 $(8^+,9^+)$&  $8^+_6$ & 1st FNU	&2197.6(4)& 6.40(14)	& 7.297 &7.297  & 19.322 & 6.877  &6.883  \\ \T\B
 $(8^+,9^+)$&  $9^+_5$ &1st FNU 	& 2197.6(4)&6.40(14)	&8.372& 9.657 & 19.322 &1.995&0.455
   
\end{tabular}
\end{ruledtabular}
\end{table*}

Recently, a new beta-decaying isomeric state (8$^-$) for $^{214}$Bi$^m$ is predicted at the CERN-ISOLDE facility, which decays to various excited states of $^{214}$Po listed in table \ref{tab:iso1} with a half-life of 9.39(10) min. In Ref. \cite{iso}, experimental energy spectra are compared with the shell-model, but shell-model results for beta-decay properties such as $\log ft$ are not reported. We here give theoretical estimates for these results. Table \ref{tab:iso1} shows the $\log ft$ and average shape factor for $^{214}$Bi((8$^-$)) isomer. Computations have been carried out for various transitions from $^{214}$Bi((8$^-$)) isomer to different excited states of $^{214}$Po by using theoretical Q-values. All the transitions in this table are first forbidden non-unique transitions except one transition that is first forbidden unique transition. Thus, we have compared our shell-model results according to the decay mode with the experimental data available in the reference mentioned above. Good results are obtained for the shell-model calculations. For instance, at excitation energy 1584.4(4) keV for a transition from $^{214}$Bi((8$^-$)) to $^{214}$Po((8$^+_1$)), experimental $\log ft$ value is 6.25(12) and the theoretical value is 6.264 for the set I and 6.255 for the set II.  We have computed $\log ft$ values for all the transitions where more than one spin-parity assignment is possible. Based on shell-model results it is possible to predict one suitable spin-parity for the state at that energy level. It can be inferred that the spin-parity for the state at energy 1633.5(4) keV can be 8$^+_2$ because its theoretical $\log ft$ value is 6.466 for the set I and 6.518 for the set II, which is close to the experimental value, i.e., 6.62(14). Also, at energy level 1843.0(4) keV, experimental $\log ft$ will be decided based on whether the decay is unique first forbidden or non-unique first forbidden beta-decay. As a result, the state is predicted to be 7$^+_1$ out of two possible spin parity states, 6$^+_1$ and 7$^+_1$, because the calculated $\log ft$ value of 7$^+_1$ state is close to the experimental value as compared to that for 6$^+_1$. At energy level 1969.1(4) keV, it is difficult to assign a unique spin-parity for the state out of two 8$^+_4$ and 9$^+_2$ states, because both spin-parity assignments give good results. At energy 2059.5(4) keV, 8$^+_5$ can be suitable as its $\log ft$ value for the set I is 6.771 and for the set II is 7.558, whereas its experimental value is 6.04(13). Also, at energy 2197.6(4) keV, 8$^+_6$ state can be predicted because its theoretical $\log ft$ value is 7.297 and experimental value is 6.40(14).

\section{Conclusion}

In the present work, we have calculated various beta-decay properties such as $\log ft$ values, shape factors, and half-lives of nuclei in the north-east region of $^{208}$Pb nucleus. These computations have been performed for $^{210-215}$Pb $\rightarrow$  $^{210-215}$Bi and $^{210-215}$Bi $\rightarrow$ $^{210-215}$Po transitions using KHPE effective interaction without any truncation. First, we have computed the quenching factor using the chi-square fitting method. Then, using this value of the quenching factor and another set taken from Ref. \cite{zhi}, we have calculated $\log ft$, shape factor, and half-lives of these transitions. We have also calculated shell-model Q-values of the concerned nuclei, and used both experimental and theoretical Q-values to study these beta-decay properties for better comparison. Our results for the $\log ft$ values and half-lives obtained by the shell-model calculations show good agreement with the available experimental data.  Use of experimental Q-values and inclusion of the quenching in both the axial and vector couplings is generally better compared with the case of theoretical Q-values and the quenching in the axial coupling only. We have also calculated $\log ft$ values for $^{214}$Bi$^m((8^-))$ isomer recently predicted at the CERN-ISOLDE facility. We have used shell-model Q-values for these calculations. We have also confirmed spins and parities of various states with the help of shell-model calculations, where unique assignments of spin-parity are not possible experimentally. Also, we have calculated $\log ft$ values of the transitions whose experimental data is not available. These shell-model results will add more information to the present data and will be very helpful for future experiments. Further, shell-model results can be improved with the inclusion of core polarization effects \cite{warburton90} by allowing nucleonic excitation across $^{208}$Pb core and tuning effective interactions.

\section*{Acknowledgement}

S. S. would like to thank CSIR-HRDG (India) for the financial support for her Ph.D. thesis work. P. C. Srivastava acknowledges a research grant from SERB (India), CRG/2019/000556. T. S. acknowledges JSPS (Japan) for a grant JSPS KAKENHI, No. JP19K03855.

\end{document}